# Visual cryptography in single-pixel imaging

SHUMING JIAO,[1] JUN FENG,[1] YANG GAO,[1] TING LEI,[1,2] XIAOCONG YUAN[1,3]

[1]*Nanophotonics Research Center, Shenzhen University, Shenzhen, Guangdong, China*
[2]*leiting@szu.edu.cn*
[3]*xcyuan@szu.edu.cn*

**Abstract:** Two novel visual cryptography (VC) schemes are proposed by combining VC with single-pixel imaging (SPI) for the first time. It is pointed out that the overlapping of visual key images in VC is similar to the superposition of pixel intensities by a single-pixel detector in SPI. In the first scheme, QR-code VC is designed by using opaque sheets instead of transparent sheets. The secret image can be recovered when identical illumination patterns are projected onto multiple visual key images and a single detector is used to record the total light intensities. In the second scheme, the secret image is shared by multiple illumination pattern sequences and it can be recovered when the visual key patterns are projected onto identical items. The application of VC can be extended to more diversified scenarios by our proposed schemes.

## 1. Introduction

Visual cryptography (VC) [1-12] is a security technique for encrypting an image in a way that the original image can be visually decrypted. In VC, a secret image is randomly expanded to multiple visual key images and each visual key image is referred to as a share. Conventionally these visual key images are printed on transparent sheets, the secret image can be visually decoded when a qualified subset of the keys are overlapped. An example of VC with two visual keys is shown in Fig. 1. In VC, from one individual visual key, no information about the secret image can be extracted. After the multiple visual keys are overlapped, the secret image can be displayed but each individual key cannot be directly observed from the overlapped result. For different security applications, both the visual keys and the secret image can be protected as invisible to unauthorized users.

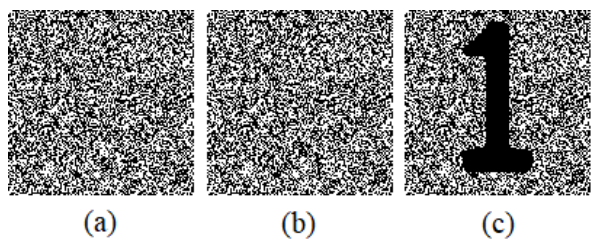

Fig. 1. Example of VC: (a)visual key 1; (b)visual key 2; (c)visually decoded secret image when the two visual keys printed on transparent sheets are overlapped.

VC can be categorized as an optical encryption technique [13-15] and the decryption of a secret image is realized optically. In comparison, both the encryption and decryption steps in conventional digital encryption schemes are implemented with computer algorithms. In the original design of VC [1,2], the visual key images are random binary patterns without carrying other meaningful information. Improved VC schemes [3-8] allow each visual key image to look like natural binary images, grayscale images or color images. In some recent works [9, 10], each visual key is designed in the form of a Quick-Response (QR) code pattern.

In VC, the visual keys are usually printed on transparent sheets. If they are printed on opaque materials (e.g. a paper sheet), it is hard to decode the secret image by overlapping. It is proposed that the visual key information can be physically embedded in holographic optical elements (HOE) in previous works [11,12]. Based on optical coherent diffraction and volume grating, the secret image can be visually displayed when the correct HOEs are placed in the optical setup and illuminated by laser. This scheme has some unique advantages compared with conventional transparency-based visual cryptography. However, the optical setup has certain experimental complexity such as precise alignment requirement of optical elements, fabrication difficulty of HOE and availability of laser source.

Different from the coherent imaging system [11,12], an incoherent optical system such as a single-pixel imaging (SPI) system [16-21] will have significantly lower experimental complexity. In fact, this has been revealed in some previous works about optical computing [22] and holography [23]. The object image is usually captured by a pixelated sensor array in a conventional optical imaging system. However, in SPI, the sensor only has one single pixel and it will collect the total light intensity of the entire object scene. The object image will be sequentially illuminated by different structured light intensity patterns and the total light intensity for each illumination is recorded by the single-pixel detector. Finally, the object image can be computationally reconstructed from the illumination patterns and the recorded single-pixel intensity sequences by various kinds of algorithms [24]. Some recent works also demonstrate that the image reconstruction can be possibly realized full-optically [25].

In VC, a key step is the overlapping of pixels from multiple transparent visual key images. In SPI, the light intensities of different pixels are superposed in the recorded light intensities by the single-pixel detector. These two schemes have common features from this perspective. In previous works [26-31], optical encryption with SPI has been extensively investigated but the VC framework has never been attempted.

In this work, VC is proposed to be combined with SPI for the first time and two novel schemes are proposed. First, it is demonstrated that a secret image recovery can be implemented from multiple QR-code visual keys printed on opaque sheets. The visual keys are overlapped in the recorded single-pixel light intensities and the secret image can be reconstructed. Each individual QR-code visual key image is not directly captured by the detector. In addition, the visual key images disguised as readable QR codes have natural visual meaning and it is less noticeable by the attackers. Second, it is proposed that the visual keys can be embedded in the illumination patterns of SPI, instead of a transparent or opaque sheet. When multiple sets of illumination patterns are projected onto identical object images and only one single-pixel detector is used to collect the total light intensity, the secret image can be recovered in the reconstruction result.

## 2. Proposed visual cryptography (VC) schemes in single-pixel imaging (SPI)

### 2.1 Principles of single-pixel imaging (SPI)

In SPI, the projection device will sequentially project N different illumination patterns $P_1(x,y)$, $P_2(x,y)$, ..., $P_N(x,y)$ onto the object image $O(x,y)$. Then a sequence of single-pixel light intensities $I_1$, $I_2$, ..., $I_N$ is recorded by the single-pixel bucket detector. For the $nth$ ($1 \leq n \leq N$) illumination pattern, $I_n$ is mathematically the inner product between $O(x,y)$ and $P_n(x,y)$, given by Equation (1).

$$I_n = \iint O(x,y) P_n(x,y) dxdy \tag{1}$$

The object image $O(x,y)$ can be reconstructed from all the illumination patterns $P_n(x,y)$ ($1 \leq n \leq N$) and the single-pixel intensity sequence $I_n$ ($1 \leq n \leq N$) by various methods [24, 25]. In this work, random binary illumination patterns are used and each pixel in $P_n(x,y)$ is randomly set to be 0 or 1. A typical optical setup for SPI is shown in Fig. 2.

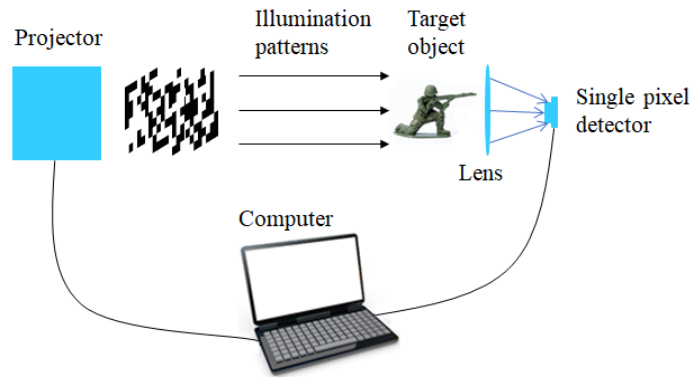

Fig. 2. Optical setup of a single-pixel imaging system.

*2.2 Proposed opaque QR-code visual cryptography with single-pixel imaging*

In our first proposed scheme, the secret image is first expanded to multiple QR-code patterns carrying the same information, which are employed as visual key images. The visual key images are printed on opaque sheets instead of conventional transparent sheets. When multiple visual key images are illuminated by multiple sets of identical illumination patterns under the recording of only one single-pixel detector, the secret image can be reconstructed from the recorded single-pixel intensity sequence.

In SPI, if the same set of illumination patterns $P_n(x, y)$ are projected to two different object images $O_1(x, y)$ and $O_2(x, y)$ in the object scene but only one single-pixel detector is used record the total light intensity, the SPI model will be given by Equation (2).

$$I_n = \iint \left[ O_1(x, y) + O_2(x, y) \right] P_n(x, y) dxdy \qquad (2)$$

The finally reconstructed image will be the overlapped result of $O_1(x, y)$ and $O_2(x, y)$, instead of each individual one. If each object image is one visual key printed on a opaque sheet in VC, a SPI system can perform the VC decoding and retrieve the secret image, shown in Fig. 3.

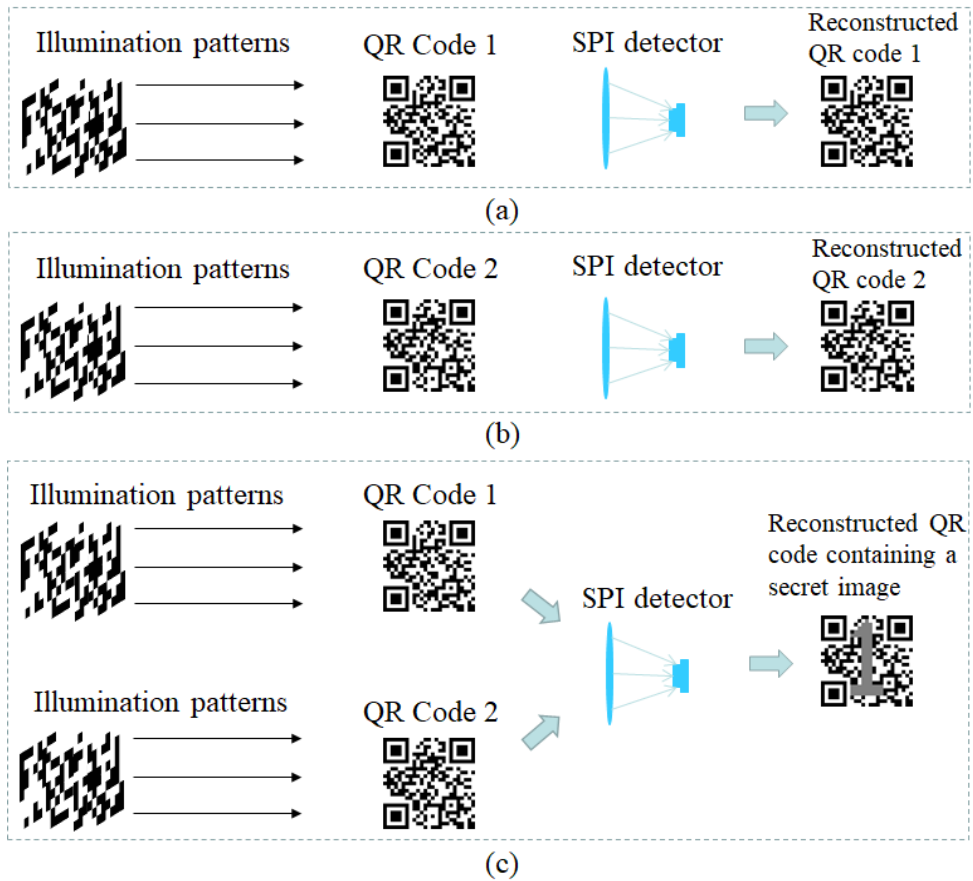

Fig. 3. QR-code-based opaque VC with SPI: (a) the first QR-code visual key is individually recorded and reconstructed; (b) the second QR-code visual key is individually recorded and reconstructed; (c) the two QR-code visual keys are simultaneously recorded by a single detector and the secret information "1" can be reconstructed.

In the previous works [11,12], visual key images are embellished into QR-code-like appearance. But they are not true QR code patterns and no information can be retrieved from them by a QR code reader. In this work, each visual key is embedded in a true QR code pattern. The advantage of employing a QR code pattern as a visual key compared with a random binary pattern is that the visual key image will have natural visual meaning and becomes less noticeable by the attackers.

One QR code pattern is essentially a square dot matrix and the intensity of each dot (black or white) can represent a binary value (0 or 1). Various kinds of information such as texts and weblinks can be stored in a QR code. The stored information in a QR code can be easily retrieved by a reading device (e.g. a smart cellphone). Due to the error correction coding mechanism, if some square dots in the pattern (within a certain percentage) are modified, the original information can still be correctly retrieved [32-34].

Each square dot has two possible intensity values 0 and 1. When two corresponding square dots in the two QR codes are superposed, there are three possible values 0, 1 and 2, shown in Table 1. If the intensity value after overlapping is 1, there is no way to determine whether it is 1 in the first dot and 0 in the second dot, or it is 0 in the first dot and 1 in the second dot, which has strong security features.

Table 1. Superposed square dot intensity values when two QR codes printed on opaque sheets are overlapped

| Square dot intensity in QR code 1 | Square dot intensity in QR code 2 | Superposed intensity |
| --- | --- | --- |
| 0 | 0 | 0 (black) |
| 1 | 1 | 2 (white) |
| 0 | 1 | 1 (gray) |
| 1 | 0 | 1 (gray) |

For a binary square dot pattern (secret image) hidden in two QR codes by VC, the encoding is realized in the following manner. It is assumed the secret binary image has the same resolution as the two QR codes. Originally the two visual keys (two QR codes) are identical and carry the same information. If one square dot in the secret image represents 0, the square dots at the corresponding positions in the two QR codes remain unchanged. If one square dot in the secret image represents 1, the square dot at the corresponding position in one of the two QR codes (randomly chosen) will be encoded as 1 and the square dot in another QR code will be encoded as 0. After encoding, when the two QR-code visual keys are overlapped, the secret image can be visually displayed. At the same time, each QR-code visual key still carries its original readable information due to the error correction coding mechanism stated above.

## 2.3 Proposed visual cryptography in the binary random illumination patterns of single-pixel imaging

Apart from being printed on transparent sheets or opaque sheets, the visual keys in VC can be embedded in the illumination patterns of SPI. Conventionally, there is only one sequence of illumination patterns $P_n(x,y)$ $(1 \leq n \leq N)$ in SPI. To implement VC, two different sets of illumination patterns $P_n(x,y)$ $(1 \leq n \leq N)$ and $Q_k(x,y)$ $(1 \leq k \leq N)$ are used. Originally, two sequences are identical and every pixel in each illumination pattern has a random binary intensity value. Similar to the encoding scheme proposed in Section 2.2, the corresponding pixels will be encoded as one "1" and one "0" for each pair of illumination patterns $(n = k)$ if one pixel in the secret image has an intensity of "1". No change will be made if one pixel in the secret image has an intensity of "0". Each individual set of illumination patterns can be used for capturing an object image in conventional SPI. However, when the two sets of patterns are simultaneously projected onto two identical object images, its equivalent that the object is illuminated by a set of virtually superposed illumination patterns containing the secret image, given by Equation (3). It shall be noted that the illumination patterns are virtually overlapped but not actually superposed in all our experiments. In the regions of each virtually superposed pattern corresponding to the foreground of the secret image, all the pixels have identical intensity values and there is no spatial variation. Consequently, these regions will be missing in the reconstructed image based on the working mechanism of SPI. Finally, the secret image can be revealed in the reconstruction result, shown in Fig. 4. In the reconstruction of secret image, either illumination pattern sequence can be used.

$$I_n = \iint O(x,y)\left[P_n(x,y)+Q_n(x,y)\right]dxdy \qquad (3)$$

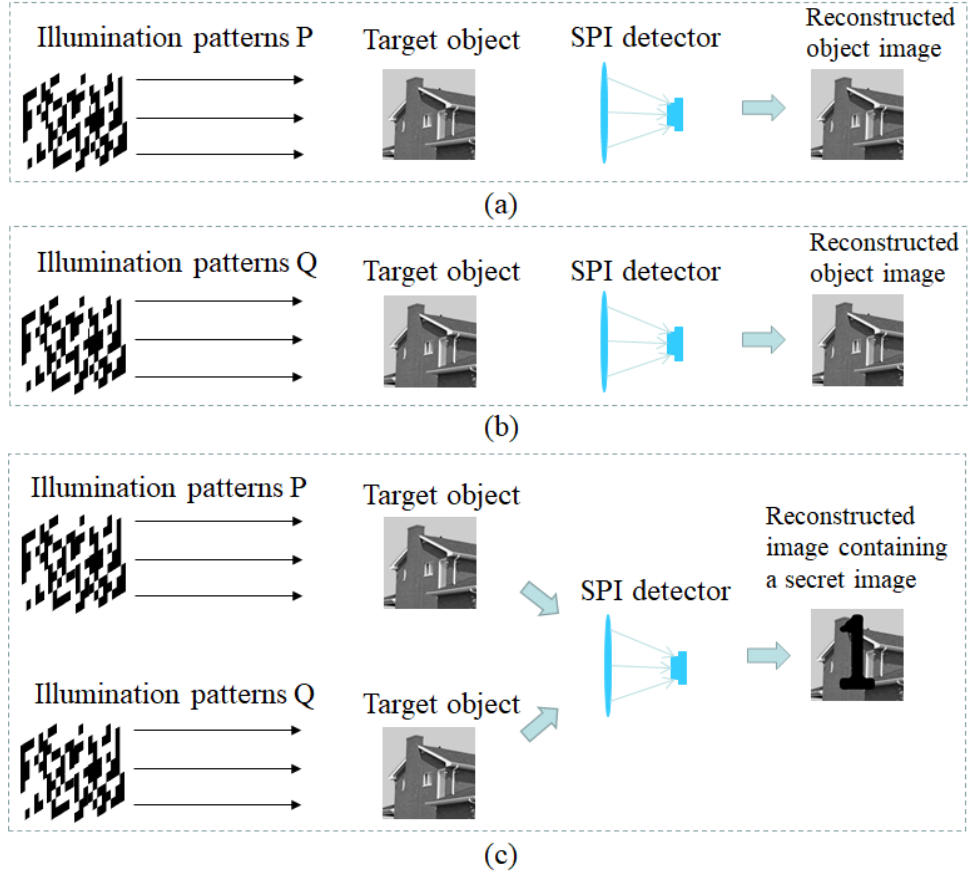

Fig. 4. Visual cryptography in the binary illumination patterns of SPI: (a) the object image is individually illuminated by the first set of visual-key illumination patterns; (b) the object image is individually illuminated by the second set of visual-key illumination patterns; (c) the two identical object images are simultaneously illuminated by the two sets of visual-key illumination patterns under a single detector and the secret information "1" can be reconstructed.

## 3. Results and discussions

The two proposed schemes in this work are verified by simulation and experimental results. The experimental setup is shown in Fig. 5. The object image is printed on a paper card and the illumination patterns are projected by a JmGO G3 projector. The single-pixel light intensities are recorded by a Thorlabs FDS1010 photodiode detector and then the data is collected by a NI USB-6216 data acquisition card.

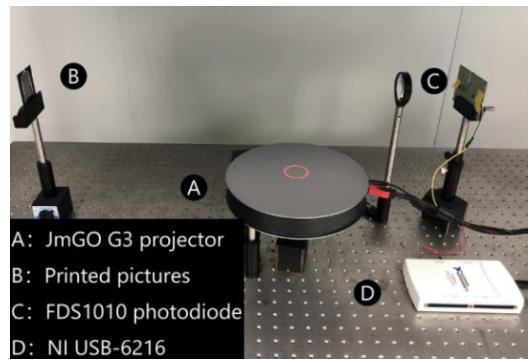

Fig. 5. Experimental setup of our single-pixel imaging system.

First, the proposed opaque QR-code VC with SPI is verified. A $33 \times 33$ square-dot QR code pattern is generated by a open-source program [35] and it contains the text information "Nanophotonics Research Center". The error correction level in the QR code generation is set to be "H" level. The text can be retrieved by any QR code reader from the QR code. Then a secret image containing the information "OK" is encoded as two QR-code visual keys by the method described in Section 2.2, shown in Fig. 6. The original information "Nanophotonics Research Center" can still be retrieved from these two modified QR codes in Fig. 6(c) and Fig. 6(d). At the same time, the secret image cannot be directly observed from each individual visual key.

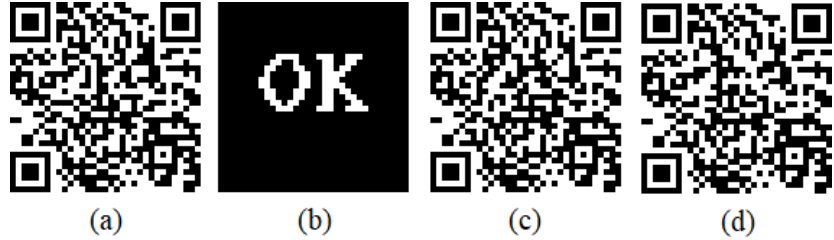

Fig. 6. (a)Original QR code containing the information "Nanophotonics Research Center"; (b)Secret image; (c)QR-code visual key 1; (d)QR-code visual key 2.

The two printed QR-code visual keys are employed as the target objects in the SPI experiment. One sequence of binary random illumination patterns with a resolution of $33 \times 33$ pixels are projected onto the target objects. The number of illumination patterns is $N = 2178$. In the optical experiment, the QR-code visual keys are printed on opaque paper cards. First, QR-code visual key 1 is individually captured and reconstructed. Then, QR-code visual key 2 is individually captured and reconstructed. Finally, both two QR-code visual key are placed nearby in the object scene and they are illuminated by two set of identical illumination patterns. The total light intensity is recorded by a single detector. The image reconstruction from the single-pixel intensity sequence is realized by a total variation minimization algorithm [24]. The reconstruction results in the simulation and optical experiment are shown in Fig. 7.

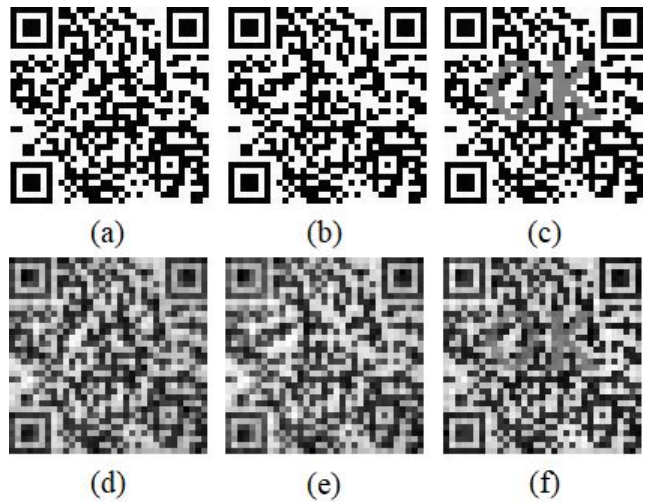

Fig. 7. Simulated reconstruction results of SPI: (a) QR-code visual key 1; (b) QR-code visual key 2; (c) both visual keys are captured by a single detector. Optical experimental reconstruction results of SPI: (d) QR-code visual key 1; (e) QR-code visual key 2; (f) both visual keys are captured by a single detector.

It can be observed that the binary secret image "OK" can be recovered when the two visual keys are illuminated by identical sequences of illumination patterns simultaneously and the light intensity is captured by a single detector. The reconstruction results in the optical experiments suffer from some noise contamination compared with the reconstruction results in the simulation. But the experimental results generally agree with the theoretical analysis and simulation results. The original information "Nanophotonics Research Center" can be retrieved from all the experimentally reconstructed QR code images by a reader even under noisy conditions, due to the error correction coding mechanism in QR coding, as stated in Section 2.

Our proposed VC in the binary random illumination patterns of SPI is verified as well. Two sets of random binary patterns containing the secret information "OK" are generated. Each pattern has a resolution of $37 \times 37$ pixels and there are totally 2738 patterns in each sequence. The first three patterns in each set are shown in Fig.8 as examples.

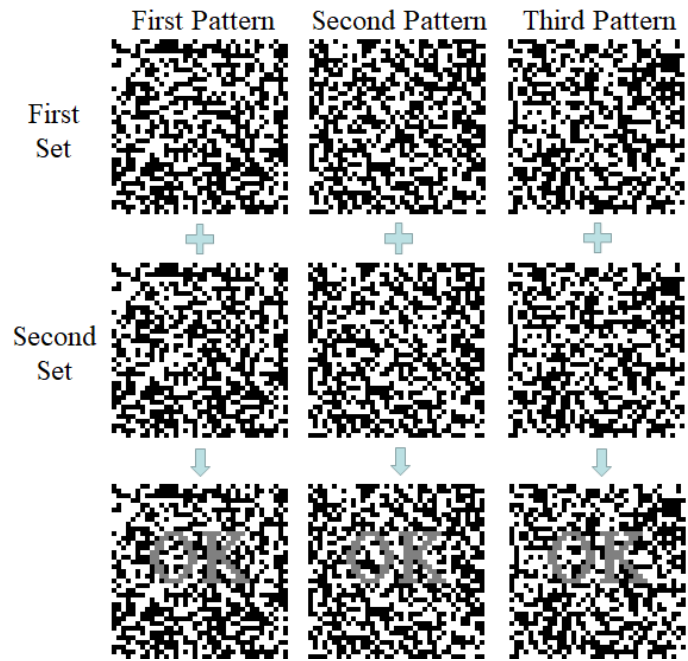

Fig. 8. First three patterns in the two sets of illumination patterns and corresponding superposed results.

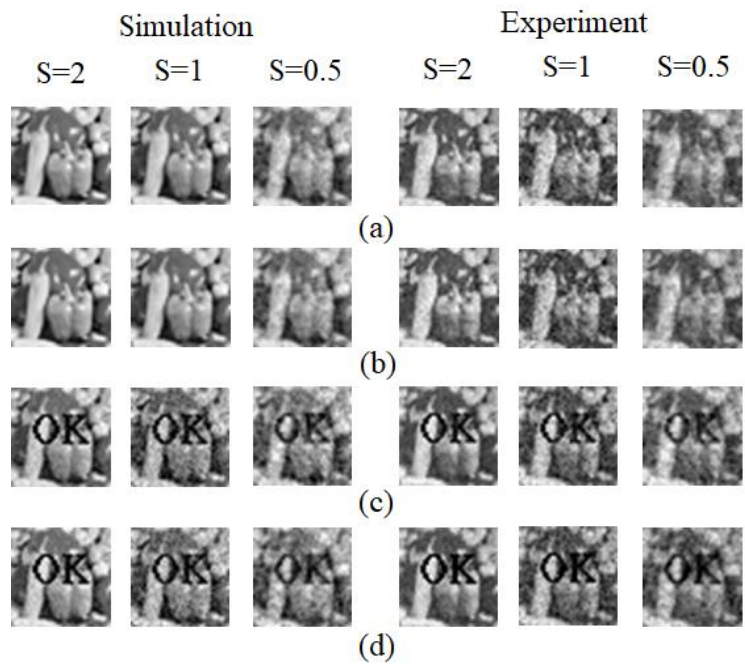

Fig. 9. Reconstruction results for our proposed VC in the binary random illumination patterns of SPI: (a) only the first set of illumination patterns are projected onto the object; (b) only the second set of illumination patterns are projected onto the object; (c) two sets of illumination patterns are projected onto two identical objects and the first set of patterns are used for reconstruction; (d) two sets of illumination patterns are projected onto two identical objects and the second set of patterns are used for reconstruction.

From each individual set of illumination patterns, the secret image is not observable. But the secret information will be recovered whenever any two corresponding patterns from the two sets are superposed. In the SPI experiment, first each illumination pattern sequence is individually projected onto a "pepper" image printed on a paper card and the object image can be reconstructed by total variation minimization from the recorded single-pixel intensities, which is a conventional SPI process. Then, two copies of the same printed "pepper" image are placed in the object scene and they are separately illuminated by each set of illumination patterns. Only one single-pixel detector is used to record the total light intensity. The object image is reconstructed based on the single-pixel intensity sequence and either set of illumination patterns. Three different sampling ratios $S = 2$, $S = 1$, and $S = 0.5$ corresponding to $N = 2738$, $N = 1369$, and $N = 595$ are attempted. The reconstruction results are shown in Fig. 9.

It can be observed that the object image can be reconstructed in a conventional way when either illumination sequence is individually projected. The quality of reconstructed images will be enhanced as the sampling ratio is higher. The experimental reconstruction results have some quality degradation compared with the simulation results due to noise and distortion. When the two illumination pattern sequences are projected onto two identical objects simultaneously, the secret image "OK" can be recovered in the reconstruction results. Even when the sampling ratio is reduced to 0.5, the secret image can still be observed. In the reconstruction, either illumination pattern sequence can be used and the visual qualities of recovered secret images are similar.

In addition, another different object image is tested under the same conditions and the results are shown in Fig. 10. It can be observed that the secret information "OK" can be recovered as well. Since the secret image is hidden in the two sets of illumination patterns, this scheme can be implemented for different object images as long as the designed illumination patterns are used.

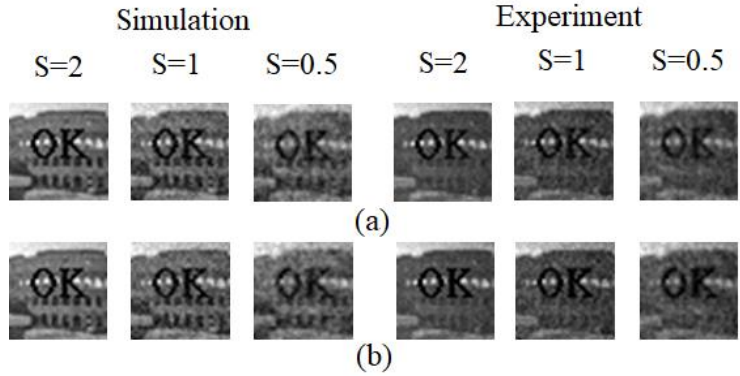

Fig. 10. Recovered secret image for our proposed VC in the binary random illumination patterns of SPI when a different object image is used: (a) reconstruction with the first set of illumination patterns; (b) reconstruction with the second set of illumination patterns.

In the current optical setup for these two visual cryptography schemes, the two object images are placed nearby and the two sets of illumination patterns are projected close to each other. In fact, the pattern illumination for the two objects are not necessarily performed nearby [27, 30]. They can be separately performed out of the line-of-sight and even at two far-away locations. The total light intensity can be collected by the single-pixel detector as long as the two object scenes are within the detection range. In comparison, the visual keys have to be placed in the same location and overlapped for decryption in conventional VC. The potential advantage of long-distance VC in our proposed scheme will be favorable for some security applications.

## 4. Conclusion

Visual cryptography (VC) is an optical image encryption technique allowing the secret image to be recovered when multiple visual key images are overlapped. Conventionally, the visual key images are printed on transparent sheets and they have to be placed at the same location for overlapping. In addition, each visual key image may be easily noticed by attackers.

In this work, VC is implemented under the framework of single-pixel imaging (SPI) and two new schemes are proposed. The overlapping of pixel intensities in VC is realized by the light intensity superposition under single-pixel detection in SPI.

In the first proposed scheme, the visual key images are printed on opaque sheets instead of conventional transparent sheets. When multiple visual key images are illuminated by multiple sets of identical illumination patterns under the recording of only one single-pixel detector, the secret image can be reconstructed from the recorded single-pixel intensity sequence. In addition, secure visual keys disguised as readable QR code patterns are designed.

In the second proposed scheme, the visual key images are embedded in the binary random illumination patterns of SPI, which is hard to be noticed by attackers. When multiple visual-key illumination patterns are projected onto multiple identical items under the recording of only one single-pixel detector, the secret image can be reconstructed as well. From each individual illumination pattern sequence, the secret image is not observable.

These two proposed schemes are verified by simulation and experimental results. The application of VC can be extended to more diversified scenarios by our proposed schemes. For example, the multiple visual keys are not necessarily to be placed at the same location for overlapping in SPI.

**Funding**

National Natural Science Foundation of China (61805145, 11774240); Leading Talents Program of Guangdong Province (00201505); Natural Science Foundation of Guangdong Province (2016A030312010).

**Disclosures**

The authors declare no conflicts of interest.